\documentclass[reprint,aip,rsi,amsmath,amssymb,nofootinbib]{revtex4-1}

\usepackage[normalem]{ulem}
\usepackage{graphicx}
\usepackage{color,soul}
\usepackage{subfigure}
\usepackage{dcolumn}
\usepackage{bm}
\usepackage{hyperref}
\usepackage{url}
\usepackage{multirow}
\usepackage{mathtools}
\usepackage{csvsimple}
\hypersetup{
    colorlinks=true,
    linkcolor=blue,
    filecolor=blue,      
    urlcolor=blue,
    citecolor=blue,
}
\usepackage{float, multirow}

\newcommand{\eg}{\textit{e.g., }}

\usepackage{longtable}

\appdef \turnpage {%
  \AddToHookNext{shipout/after}{%
    \global\pdfpageattr\expandafter{\the\pdfpageattr/Rotate 90}%
    \AddToHookNext{shipout/after}{%
      \global\pdfpageattr\expandafter{\the\pdfpageattr/Rotate 0}%
    }%
  }%
}

\begin{document}

\title{Beam Sources for 10 TeV Wakefield Collider}

\author{Oksana Chubenko}
\email[E-mail address: ]{chubenko@niu.edu}
\affiliation{Department of Physics, Northern Illinois University, DeKalb, IL 60115, USA}

\author{Siddharth Karkare}
\affiliation{Department of Physics, Arizona State University, Tempe, AZ 85287, USA}

\author{Joe Grames}
\affiliation{Center for Injector and Sources, Thomas Jefferson National Accelerator facility, Newport News, VA 23606, USA}

\author{Matthias Fuchs}
\affiliation{Karlsruhe Institute of Technology, Germany}

\begin{abstract}
Due to its unique advantages, wakefield particle acceleration has been proposed as a promising pathway toward a 10 TeV collider. Several concepts, including Laser Wakefield Acceleration (LWFA), Plasma Wakefield Acceleration (PWFA), and Structure Wakefield Acceleration (SWFA), are being actively explored as potential approaches toward a 10 TeV collider. Each of these approaches requires particle sources (for the witness beam or for both drive and witness beams) with specific parameter sets to enable efficient wakefield acceleration. This work represents evaluation of existing and emerging particle generation technologies in the context of specific 10 TeV wakefield collider design requirements, with a particular focus on achievable brightness. 

\end{abstract}
\maketitle

\section{\label{sec:introduction}Introduction}

Wakefield acceleration is a novel advanced particle acceleration technique in which intense laser pulses or charged-particle beams excite strong electromagnetic wakefields in a plasma or engineered structure. These wakefields can sustain accelerating gradients that are orders of magnitude higher than those achievable with conventional radio-frequency (RF) accelerators. As a result, wakefield accelerators have the potential to significantly reduce the size and cost of future accelerator facilities.

In addition to their compactness, wakefield acceleration schemes offer a promising path toward multi-TeV particle colliders and next-generation light sources. Their ability to support ultra-high accelerating gradients makes them attractive candidates for extending the energy frontier beyond the practical limits of conventional RF technology. Therefore, the wakefield particle acceleration has been suggested~\cite{Gessner_2025} as a promising approach toward a 10 TeV collider.

Particle sources are a critical component of any future collider, as its conceptual design and overall performance strongly depend on the quality of the generated beams. Wakefield acceleration schemes require electron and/or positron beams with exceptionally high brightness, low emittance, high stability, and, in many cases, high bunch charge and repetition rate. High spin polarization is also essential for colliders that rely on spin-polarized particle beams. The source parameters directly influence beam capture, acceleration efficiency, luminosity, and overall collider feasibility.

Because different wakefield concepts impose different requirements on drive and witness beams, the development of advanced particle sources capable of meeting these demanding specifications remains one of the key challenges for future collider realization. Consequently, systematic evaluation and optimization of existing and emerging source technologies are essential for the successful implementation of a 10 TeV wakefield collider. This work is focused on the assessment of capabilities of a wide range of particle generation techniques in the context of specific collider design requirements, with a particular focus on achievable brightness.

\section{\label{sec:requirements}Beam Requirements}

Several wakefield acceleration schemes are currently being explored as potential candidates for the future 10 TeV wakefield collider. These approaches differ primarily in the medium used to sustain the wakefields and in the mechanism used to drive them. Below, we briefly describe these schemes, with particular emphasis on the beam parameters required for their implementation in the context of the 10 TeV collider. A summary of the preliminary beam requirements (as of June 2025) for the witness beams or for both drive and witness beams of these schemes is provided in Table~\ref{requirements}.

\textbf{Laser Wakefield Acceleration (LWFA)}\cite{Tajima_1979, Tajima_2020} uses an ultra-intense, ultra-short laser pulse propagating through a plasma to excite strong plasma waves capable of accelerating charged particles at extremely high gradients. LWFA offers compact accelerator structures and has demonstrated multi-GeV electron acceleration over centimeter-scale distances. However, challenges remain in achieving high repetition rate, beam stability, and collider-scale efficiency.

LWFA is typically considered in two operating regimes: the \textit{bubble (or nonlinear) regime} and the \textit{quasi-linear regime}. These regimes describe basically how strongly the laser perturbs the plasma - and therefore how the wake structure, beam quality, and acceleration physics differ. The bubble regime naturally focuses electrons. Positrons would experience defocusing. Therefore, it is not suitable for positron acceleration. The quasi-linear regime is suitable for the acceleration of both electrons and positrons. Both regimes require relatively high bunch charges (approximately 1.3 nC and 0.2 nC, respectively), beam energies on the order of 1 GeV with energy spreads well below 1$\%$, and ultra-low normalized emittances (particularly in the quasi-linear regime, where values of 0.01-0.1~mm$\cdot$mrad are required)\cite{Benedetti_2025}.

\textbf{Plasma Wakefield Acceleration (PWFA)}~\cite{Litos_2014} employs a high-energy charged electron beam, rather than a laser, to drive wakefields in a plasma. In an \textit{electron-driven regime} of PWFA, a high-charge ($\geq$10~nC), ultra-short electron bunch (the drive beam) enters the plasma and sets up longitudinal plasma oscillations behind the beam. These oscillations form a plasma wakefield. If another, a few nC, unpolarized or polarized electron or positron bunch (the witness beam) is injected at the right phase of the wake, it experiences a huge accelerating field and gains energy. The particle source must have fine-tuning capability to provide various parameters such as bunch spacing and the number of bunches, crucial for optimizing the PWFA section. The scheme has demonstrated highly efficient energy transfer and excellent accelerating gradients, making it particularly attractive for future collider applications. Ongoing research focuses on preserving beam quality, staging multiple acceleration sections, and achieving state-of-the-art beam compression and stability. 

In the \textit{laser-gated regime} of PWFA, the electron beam still drives the wake, but the laser controls when and where electrons are injected. This regime allows precise control over injection phase. However, it adds complexity to the system (laser synchronization, optics, etc.). The drive beam of this regime requires a relatively novel source which could produce drive bunch train with 10-12 bunches with about 700~pC each separated by 100’s of fs\cite{Storey_2025}. This regime would work only for the acceleration of unpolarized electrons.

\textbf{Structure Wakefield Acceleration (SWFA)}\cite{Gai_1988, Jing_2022} utilizes electromagnetic wakefields generated inside engineered structures, such as dielectric or metallic waveguides, by intense charged-particle drive beams. Compared with plasma-based approaches, SWFA offers greater control over accelerator geometry and potentially improved stability and alignment tolerances. 

In the \textit{two-beam acceleration regime}, the drive beam and witness beam propagate in separate structures. The energy extracted from the drive beam is transferred through the wakefields to a separate trailing witness beam, which is accelerated with high efficiency. To generate high accelerating gradients for the witness bunch, very high drive-beam charge (on the order of $\sim$50 nC) is required. Typical target normalized emittances for the drive beam are from 10 to 100~mm$\cdot$mrad. The witness beam, in turn, requires a moderate bunch charge ($\sim$0.5 nC) and a low emittance ($\textless$ 1~mm$\cdot$mrad) to achieve high brightness\cite{Jing_2025}. This regime is suitable for accelerating both electrons and positrons, including polarized and unpolarized beams.

\onecolumngrid

\begin{table}[!b]
\caption{Summary of preliminary beam requirements (as of June 2025) for various wakefield acceleration schemes. The spin-polarization capability of electron ($e^-$) or positron ($e^+$) beams is defined by the upward arrow ($\uparrow$). A question mark (?) defines uncertain parameters.}
\centering
\begin{tabular}{c}
\includegraphics[width=1.\textwidth,clip]{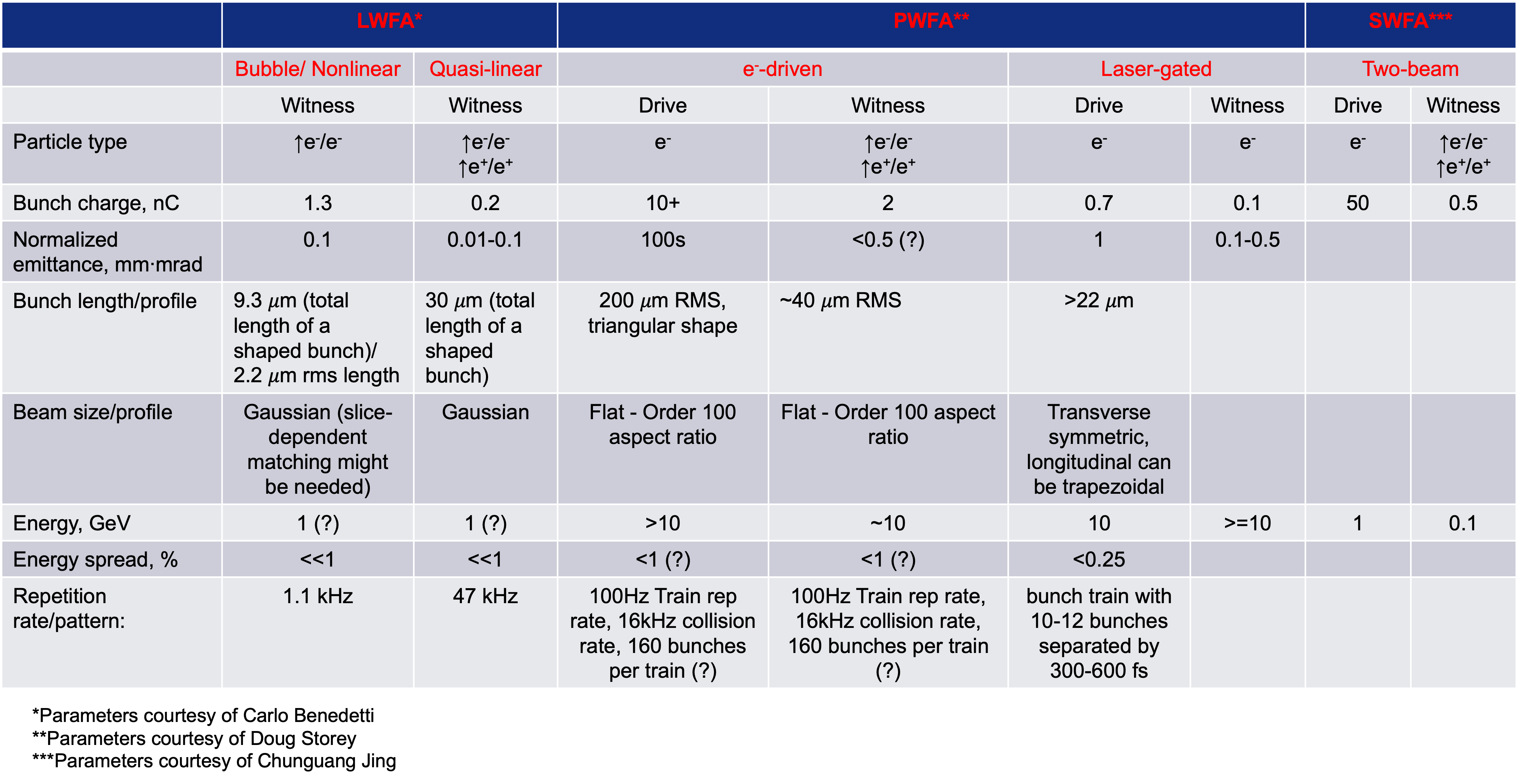}
\end{tabular}
\label{requirements}
\end{table} 

\twocolumngrid

\section{\label{sec:electrons} Electron Sources}

Pulsed electron beams are typically generated through photoemission from light-sensitive materials known as photocathodes, which may be metallic or semiconducting. When a photocathode is illuminated by pulsed laser light of sufficiently high photon energy, it emits electron bunches. An external electric field is then required to accelerate the emitted electrons and form a collimated beam.

Depending on how the accelerating field is applied, two primary technologies are commonly used: DC photoguns and RF photoinjectors\cite{Hernandez_2008}. In a DC photogun, electrons are generated inside a vacuum chamber where a static electric field is applied between the photocathode and the anode. In contrast, an RF photoinjector uses a specially shaped RF cavity containing the photocathode, where time-varying electromagnetic fields accelerate the emitted electrons to relativistic energies. In both approaches, the beam can subsequently be accelerated in downstream RF cavities, while additional solenoids and RF structures provide emittance compensation and bunch compression when required.

A third approach is based on superconducting RF (SRF) technology. SRF photoinjectors operate similarly to conventional RF photoinjectors but employ superconducting niobium cavities maintained at cryogenic temperatures, enabling high-duty-cycle or continuous-wave operation with reduced RF losses.

These conventional electron source technologies are mature, reliable, and widely used in accelerator facilities. However, their implementation in future wakefield colliders may require complex injection systems for coupling beams into plasma stages and can involve substantial cost and infrastructure requirements.

An alternative and rapidly developing approach relies on the \textit{in situ} generation of electrons directly within the plasma. This concept offers a potentially simpler and more cost-effective solution, although it remains at a comparatively early stage of development. A broader review of plasma-based particle sources can be found in Ref.~\onlinecite{Fuchs_2024}.

So, a broad range of various operational electron generation technologies exists, and there is a growing number of new facilities in various stages of planning or active construction. These technologies include both traditional and plasma-based production of polarized and unpolarized electrons. In the following sub-sections, we review several of the most prominent electron source technologies relevant to this study. Selected parameters reported in the literature for these technologies are summarized in Table~\ref{dem_param}.

\subsection{\label{subsec:pol_electrons} Polarized Electron Sources}

Traditionally, spin-polarized electron beams for applications in nuclear and high-energy physics are generated through photoemission from GaAs-based superlattice (SL) structures\cite{Maruyama_2004}. These structures consist of alternating thin layers of GaAs and GaAs alloys grown epitaxially on top of one another, creating mechanical lattice strain that lifts the valence-band degeneracy. As a result, when the photocathode is illuminated with circularly polarized light whose photon energy is close to the band-gap energy, optical transitions occur preferentially from selected spin states, yielding nearly 100$\%$ initial electron spin polarization (ESP). During electron transport to the surface, however, various depolarization mechanisms slightly reduce the final polarization.

To achieve maximum ESP, photoexcitation must occur near the band-gap energy. This requires reducing the initially high electron affinity of the material to zero or negative values, thereby enabling electron emission from the bottom of the conduction band. The only established method for achieving this condition is the deposition of Cs and O$_2$ monolayers onto the photocathode surface, a process referred to as activation to negative electron affinity (NEA)\cite{Bell_Negative}.

\begin{figure}[!b]
\begin{center}
\includegraphics [width=0.45\textwidth,clip]{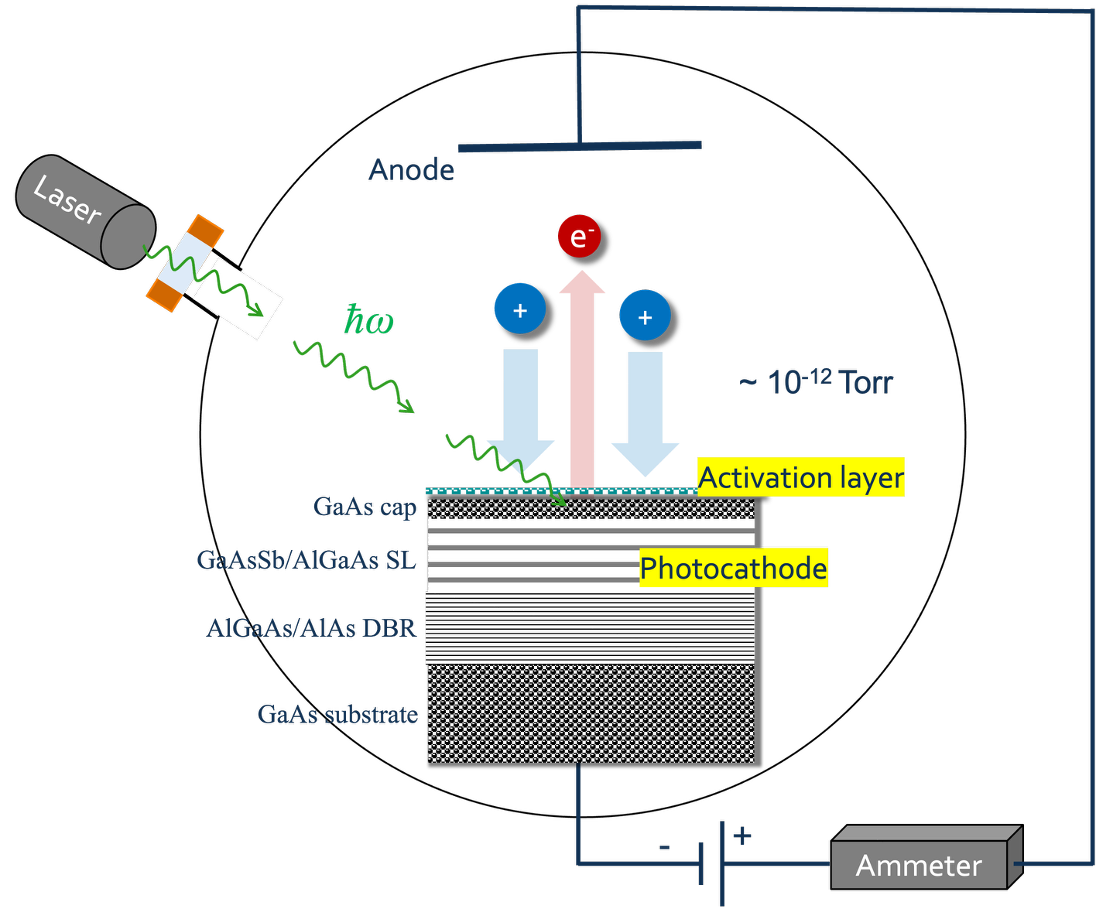}
\caption{Photoemission of spin-polarized electrons from GaAs SL structures.} 
\label{photocathode}
\end{center}
\end{figure}

Current state-of-the-art GaAs-based SL photocathodes with distributed Bragg reflectors (DBRs) can produce electron beams with ESP values of approximately 84$\%$ and quantum efficiencies (QE) of about 6.4$\%$\cite{Liu_2016}. By deploying a DBR and creating a Fabry-Perot resonating cavity, the QE can be increased up to 15\% while maintaining a high ESP of ~75$\%$~\cite{Biswas_AIP}. In principle, such combinations of high polarization and relatively high QE make these photocathodes highly attractive sources. However, the NEA activation layer is extremely fragile and highly sensitive to vacuum conditions. In particular, it degrades under ion back-bombardment, a process in which positively charged ions (generated through interactions of the emitted electron beam with residual gas molecules) are accelerated back toward the photocathode surface, damaging the activation layer (see Fig.~\ref{photocathode}). Consequently, GaAs NEA photocathodes require operation under extreme high vacuum (XHV) conditions.

The most important consequence of this requirement is that spin-polarized GaAs sources must be operated in low-gradient ($\sim$5 MV/m), low-voltage ($\sim$50–320 kV) DC electrostatic electron guns\cite{Litvinenko_2026}. Such operating conditions are necessary to reduce the risk of electric-field breakdown and to minimize dark current emission, which could otherwise degrade the vacuum, enhance ion back-bombardment, and shorten the photocathode lifetime. DC guns provide the XHV environment, at the level of 10$^{-12}$ torr, required for the spin-polarized NEA GaAs technology. However, these accelerating gradients are insufficient to bring the electrons to relativistic energies. Therefore, downstream RF accelerator cavities are used to further accelerate the beam\cite{Hernandez_2008}. Though, it should be noted that DC gun operation at voltages of 500 kV and above has been successfully demonstrated\cite{Nishimori_2014,Obina_2019}.

So, the combination of a GaAs-based high-voltage DC photogun and an (S)RF linac has become the gold standard for generating highly intense, highly spin-polarized electron beams. A representative example is the high-voltage DC gun developed at Brookhaven National Laboratory (BNL)\cite{Wang_2022}. During stable operation at 300 kV, this system demonstrated the generation of 7.5 nC bunch charges with a normalized emittance of 3.4 mm$\cdot$mrad. The typical ESP achieved with this technology is approximately 85 – 90$\%$. Additional parameters are summarized in Table~\ref{dem_param}.

Unlike electrostatic guns, RF guns are capable of sustaining substantially higher accelerating gradients and voltages in the megavolt range. Coupling these conditions with the NEA characteristics of GaAs photocathodes has the potential to produce electron beams with extremely low transverse emittance. However, past efforts to integrate GaAs photocathodes into RF guns have been largely unsuccessful due to the significantly higher dark current levels and inferior vacuum environment characteristic of RF gun operation\cite{Litvinenko_2026}.

It has recently been demonstrated at BNL\cite{Litvinenko_2026} that GaAs photocathodes can be successfully operated in a superconducting RF gun. Serving as a powerful cryogenic vacuum pump, a properly designed SRF gun operating at 4~K can provide vacuum conditions suitable for the long lifetime of GaAs photocathodes. 

The experiment was conducted using the CW CeC SRF gun system at BNL’s RHIC facility. The system was originally designed for operation with alkali antimonide photocathodes, which are considerably more robust than NEA GaAs photocathodes. The initial QE of one of the GaAs samples was 4.8$\%$; however, it decreased to approximately 0.1$\%$ during the photocathode transfer process. During subsequent operation, the QE gradually decreased to 0.04$\%$ over approximately 150 hours of operation. Despite this degradation, the photocathode successfully operated at SRF gun voltages up to 1.4 MV and generated bunch charges exceeding 1 nC at accelerating voltages near 1 MV.

Future spin-polarized SRF guns will require more reliable, and potentially simpler, photocathode transfer systems with \textit{in situ} activation capability\cite{Litvinenko_2026}. In parallel, substantial research efforts are focused on improving photocathode robustness and lifetime. These developments include more durable Cs-based activation layers\cite{Bae_2018, Cultrera_2020}, protective surface coatings\cite{Biswas_2025}, and the exploration of Cs-free spin-polarized photocathodes such as GaN\cite{Marini_2018, Cultrera_2022}.

A novel concept for generating spin-polarized electrons directly inside a beam-driven plasma wakefield accelerator has recently been proposed\cite{Nie_2021}. In this approach, a relativistic drive electron beam propagates through a gas mixture composed of, \eg Li and Xe, whose atoms possess different ionization potentials. The drive beam fully ionizes the Li atoms, thereby forming the plasma wake, while leaving the Xe atoms largely unionized. An appropriately delayed, circularly polarized ultrashort laser pulse, copropagating with the drive beam, is then focused near the entrance of the Li plasma to strong-field ionize the Xe atoms and generate spin-polarized electrons. These electrons become trapped in the wakefield potential and are subsequently accelerated to predicted energies of approximately 2.7 GeV over a distance of about 11 cm with minimal depolarization. Simulations predict the production of a high-current (0.8 kA), ultralow-emittance (37 nm normalized emittance), high-energy (2.7 GeV) electron beam with net spin polarization reaching 31$\%$.

\subsection{\label{subsec:unpol_electrons} Unpolarized Electron Sources}

For unpolarized electron beams, a significantly wider range of source technologies is available. One example is the high-voltage DC photoinjector at Cornell University, which employs a NaKSb photocathode with a typical QE$\sim 5\%$. It was demonstrated\cite{Bartnik_2015} that the injector parameters can be optimized to generate low-emittance electron beams at beam energies of 9 - 9.5 MeV. In particular, the system was shown to produce 1 nC bunches with an rms bunch length of 7 ps and a normalized emittance of 2.3 $\mu$m.

RF photoinjectors, on the other hand, generally provide substantially higher accelerating gradients, potentially lower emittances, and shorter bunch lengths. However, photocathode lifetime and reliable long-term operation remain major challenges for these systems. As an example, a high-brightness RF photoinjector is being developed for the generation of electron beams intended for subsequent acceleration in a plasma wakefield accelerator at the \text{EuPRAXIA@SPARC\_LAB} facility\cite{Silvi_2024, Del_Dotto_2025}. The photoinjector consists of an RF gun employing a low-QE Cu photocathode, followed by an X-band standing-wave section and four S-band traveling-wave accelerating structures. This system is expected to generate bunch charges in the range of 200–500 pC with normalized emittances of approximately 0.5 mm$\cdot$mrad. Additional projected parameters are summarized in Table~\ref{dem_param}.

Michigan State University (MSU), in collaboration with SLAC National Accelerator Laboratory and other partners, is developing a low-emittance superconducting SRF gun for the LCLS-II-HE project to support electron beam energies up to 8 GeV\cite{Yin_2025,Miller_2023}. A prototype SRF gun is currently under development at MSU. High-power operation has already been successfully demonstrated using a Cu cathode, while high-power testing with low-emittance photocathodes, such as CsKSb and related materials, is expected in the near future. The technology is projected to achieve normalized emittances of approximately 0.1 mm$\cdot$mrad at bunch charges of 100 pC.

\section{\label{sec:positrons} Positron Sources}

Positron beams are conventionally generated through electron–positron pair production induced by bremsstrahlung photons. In this approach, a relativistic electron beam is directed onto a high-$Z$ target, producing high-energy photons that subsequently generate electron–positron pairs.

Electron–positron pair production in plasma environments is also physically feasible and has been experimentally observed in high-intensity laser–plasma interactions. However, the controlled in situ production of high-brightness positron beams suitable for accelerator applications has not yet been demonstrated and remains at the conceptual and early research stage.

Here, we provide several state-of-the-art examples relevant to this study. A more detailed review of positron sources can be found elsewhere\cite{Chaikovska_2022}. 

\subsection{\label{subsec:pol_positrons} Polarized Positron Sources}

Polarized positron beams are a key requirement for the International Linear Collider (ILC). According to the ILC Technical Design Report\cite{Adolphsen_2013} and baseline design documentation, the undulator-based scheme remains the reference approach for producing polarized positrons.
In this scheme, the primary electron beam is first generated by the photoinjector and then transported through the damping ring, followed by bunch compression stages, before entering the main superconducting RF linac, where it is accelerated to energies of approximately 150 GeV in 3.2 nC bunches. The high-energy electron beam is then passed through a superconducting helical undulator, producing a high-intensity, multi-MeV (10–30 MeV) photon beam. These photons impinge on a thin, rotating high-$Z$ target (typically a Ti-alloy), where electron–positron pairs are created. When circularly polarized radiation from the helical undulator is preserved—potentially enhanced through photon collimation—the resulting polarization is partially transferred to the produced positrons, enabling the generation of a polarized positron beam.

The produced positrons are subsequently captured, accelerated, and transported to the positron damping ring for emittance reduction prior to injection into the main linac. In the baseline configuration, the system naturally yields longitudinal positron polarization of approximately 30$\%$, with provisions for future upgrades toward $\sim$60$\%$. A proof-of-principle demonstration (Experiment E-166 at SLAC National Accelerator Laboratory) validated polarized positron production using a helical undulator. The experiment confirmed that the method is technically feasible and scalable to collider-relevant parameters, with measurements indicating successful generation and detection of polarized positrons.

It is also worth noting that alternative approaches — such as Bremsstrahlung-based schemes using polarized electron beams (e.g., at KEK and Thomas Jefferson National Accelerator Facility) and Compton-based schemes using polarized gamma rays (e.g., at SPring-8) — have been experimentally demonstrated. However, these approaches are generally considered less suitable for high-luminosity linear collider applications.

\subsection{\label{subsec:unpol_positrons} Unpolarized Positron Sources}

The current state-of-the-art unpolarized positron source is implemented at SuperKEKB, which operates one of the highest-intensity positron sources in the world\cite{Alharthi_2025,Satoh_2016}. The facility uses a conventional scheme in which primary electrons from the injector linac are accelerated to energies of approximately 3.0–3.5 GeV and delivered in high-charge bunches of about 10 nC. These electrons are then directed onto a high-$Z$ target, typically tungsten, where electromagnetic interactions in the nuclear field produce electron–positron pairs. The generated positrons are captured and focused using a combination of a strong magnetic flux concentrator and solenoidal optics. Following capture and initial acceleration, the positrons are injected into a damping ring at approximately 1.1 GeV, where synchrotron radiation damping reduces the beam emittance by several orders of magnitude. After damping, the positron bunches are re-accelerated in the linac to around 4 GeV for injection into the SuperKEKB low-energy ring (LER).

\newpage

\onecolumngrid

\begin{table}[!b] 
\caption{Summary of beam parameters achievable with present state-of-the-art particle sources (blue) and with upcoming or predicted technologies (green).}
\centering
\begin{tabular}{c}
\includegraphics[width=1.\textwidth,clip]{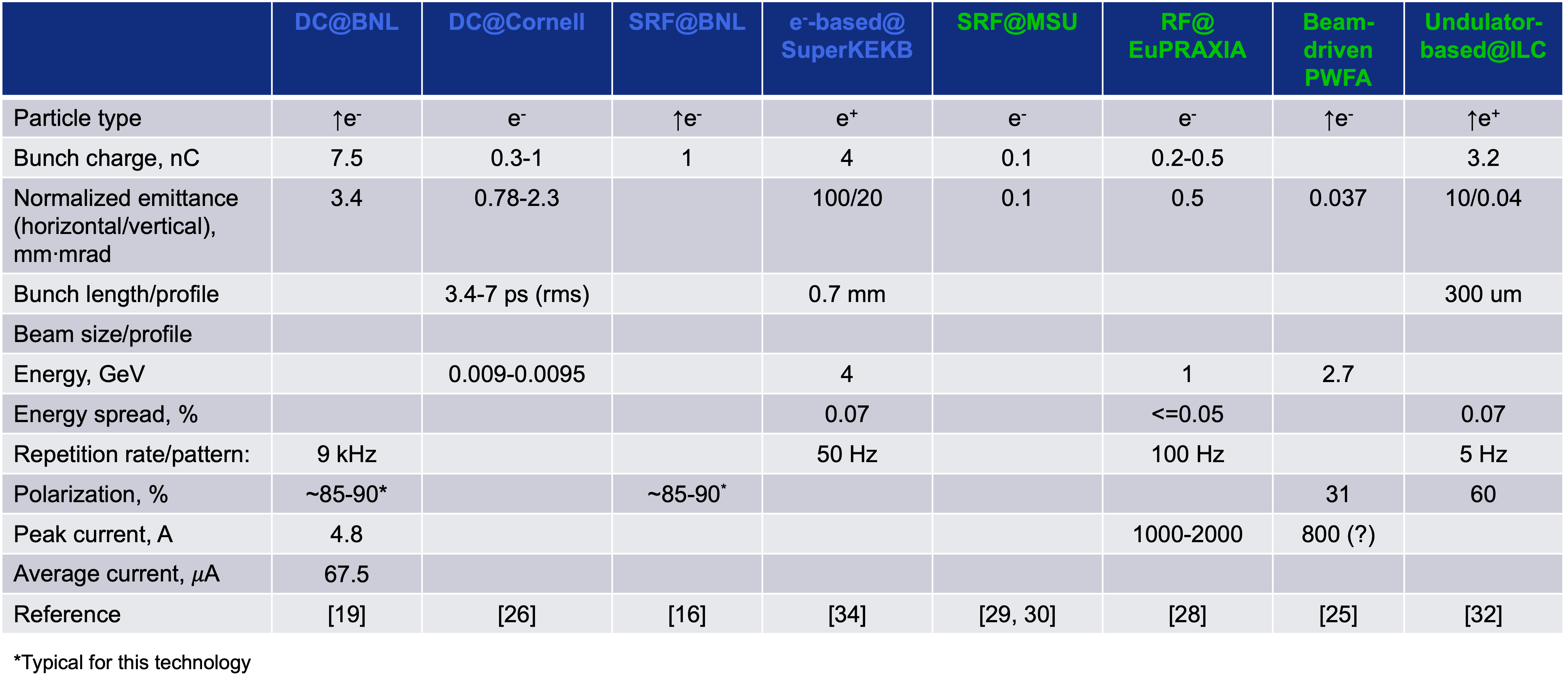}
\end{tabular}
\label{dem_param}
\end{table} 

\twocolumngrid

\section{\label{sec:results}Results and Discussion}

\begin{figure}[!b]
\begin{center}
\includegraphics [width=1\textwidth,clip]{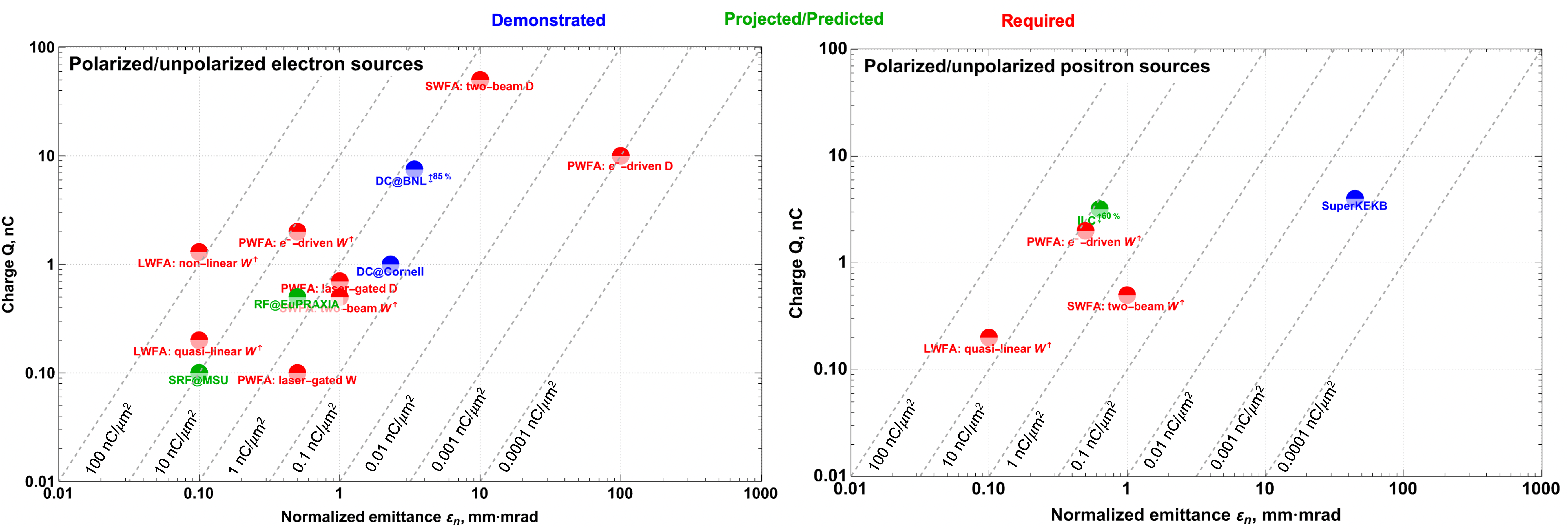}
\caption{Comparison of the electron (left) and positron (right) sources required to enable different wakefield acceleration schemes with some operational and projected state-of-the-art particle generation technologies.} 
\label{electrons_positrons}
\end{center}
\end{figure}

A comparison of capabilities of different particle generation technologies in terms of normalized emittance and bunch charge is presented in Fig.~\ref{electrons_positrons} for electrons (left) and positrons (right). Operational facilities are shown in blue, while projected or simulated capabilities are shown in green. The required source parameters for the proposed wakefield acceleration schemes are indicated in red and represent preliminary target parameters provided by the corresponding wakefield working group leaders. Spin-polarization capability is indicated by the upward arrow, $\uparrow$. The sloped lines represent contours of constant $Q/\epsilon_n^2$, as labeled, providing an indication of the achievable 4D beam brightness.

\onecolumngrid

\twocolumngrid



\section{\label{sec:conclusions}Summary}

This work represents a review of current operational state-of-the-art particle sources (polarized and unpolarized electrons and positrons), as well as emerging and projected technologies, with a focus on key performance parameters (including bunch charge and emittance) required by proposed wakefield accelerator schemes. Our preliminary analysis indicates that, in their current form, some schemes require combinations of parameters whose feasibility remains uncertain. We will continue this effort in collaboration with wakefield acceleration working groups, with increased focus on additional required beam parameters and projected cost.

\begin{acknowledgments}
This work was supported by the U.S. National Science Foundation under Award PHY-1549132, the Center for Bright Beams. This material is based upon work supported by the U.S. Department of Energy, Office of Science, Office of Nuclear Physics under contract DE-AC05-06OR23177. The authors would like to acknowledge LWFA, SWFA, and PWFA group leaders, particularly Carlo Benedetti (LBNL), Francesco Massimo (CNRS), Chunguang Jing (ANL), Xueying Lu (NIU/ANL), Alex Knetsch (SLAC), Doug Storey (SLAC), and Livio Verra (INFN), for provided lists of parameters and valuable discussions.
\end{acknowledgments}

\section*{References}
\bibliography{References}

\end{document}